%% file: paper.tex
\documentclass[letterpaper,twocolumn,10pt]{article}
\usepackage[square,comma,numbers,sort&compress]{natbib}
\usepackage{usenix}
\usepackage{times,mathptmx}
\usepackage{epsfig,amsmath,amssymb,multirow,color,xspace,fancyvrb,verbatim,url,listings,verbatim,alltt}
\usepackage{caption,subcaption}
\usepackage[T1]{fontenc}
\usepackage[breaklinks,hidelinks]{hyperref}
\usepackage{inconsolata}

\usepackage[explicit]{titlesec}

\input{glyphtounicode}
\usepackage{paralist}
\usepackage{balance}

\makeatletter
\newif\ifdraft\draftfalse
\newif\ifnotes\notestrue
\notestrue
\ifdraft\else\notesfalse\fi
\definecolor{xxxcolor}{rgb}{0.8,0,0}
\long\def\XXX{\@ifnextchar[{\@XXX}{\@XXX[]}}
\long\def\@XXX[#1]{\@ifnextchar[{\@@@XXX{#1}}{\@@XXX{#1}}}
\ifnotes
\long\def\@@XXX#1#2{{\color{xxxcolor} [XXX \ifx&#1&\else#1: \fi#2]}\xspace}
\long\def\@@@XXX#1[#2]{{\color{xxxcolor} [XXX #1: #2]}\xspace}
\else
\long\def\@@XXX#1#2{\ignorespaces}
\long\def\@@@XXX#1[#2]{\ignorespaces}
\fi

\usepackage[noend,figure]{algorithm2e}
\titleformat{\section}{\normalfont\large\bfseries}{\thesection.}{0.5em}{#1}
\titleformat{\subsection}{\normalfont\bfseries}{\thesubsection.}{0.5em}{#1}

\titlespacing{\section}{0pt}{\parskip}{\parskip}
\titlespacing{\subsection}{0pt}{5pt}{5pt}
\titlespacing{\subsubsection}{0pt}{0.5em}{0.5em}
\titlespacing{\paragraph}{0pt}{0.5em}{0.5em}

\begin{document}

\title{\vskip-0.25in\Large\bfseries The Impact of Timestamp Granularity in Optimistic Concurrency Control}

\author{
{\rm Yihe Huang\qquad Hao Bai\qquad Eddie Kohler}\\
Harvard University
\and
{\rm Barbara Liskov}\\
MIT
\and
{\rm Liuba Shrira}\\
Brandeis University
}

\maketitle

\input{abstract}
\input{intro}
\input{related}
\input{impl}
\input{eval}
\input{conclusion}

\bibliographystyle{abbrvnat}
\bibliography{occ,kpubs,kvenues}


\end{document}

%% file: abstract.tex
\section*{Abstract}

\emph{Optimistic concurrency control} (OCC) can exploit the strengths of
parallel hardware to provide excellent performance for uncontended transactions,
and is popular in high-performance in-memory databases and transactional systems.
But at high contention levels, OCC is susceptible to frequent aborts,
leading to wasted work and degraded performance.
Contention managers, mixed optimistic/pessimistic concurrency control
algorithms, and novel optimistic-inspired concurrency control
algorithms, such as TicToc~\cite{yu2016tictoc}, aim to address this problem, but
these mechanisms introduce sometimes-high overheads of their own. We show that
in real-world benchmarks, traditional OCC can outperform these alternative
mechanisms by simply adding \emph{fine-grained version timestamps} (using
different timestamps for disjoint components of each record). With
fine-grained timestamps, OCC gets $1.14\times$ TicToc's throughput
in TPC-C at 128 cores (previous work reported TicToc
having $1.8\times$ higher throughput than OCC at 80 hyperthreads). Our study shows that
timestamp granularity has a greater impact than previously thought on the
performance of transaction processing systems, and should not be overlooked
in the push for faster concurrency control schemes.

%% file: intro.tex
\section{Introduction}
\label{sec:intro}

Software running on shared-memory multi-core machines can perform and scale
excellently if it uses machine resources well~\cite{clements13scalable}.
An important design principle is to avoid extensive sharing of
frequently-written cache lines, which causes expensive locking at the
underlying cache coherence protocol level. Optimistic concurrency control
(OCC)~\cite{kung1981optimistic} obeys this principle. It avoids writing memory
for objects that are merely read, thus limiting the instances of read/write
conflicts to those that are absolutely essential for the correct operation of
the concurrency control (CC) mechanism. As a result, OCC is central to
many recent very fast transaction processing systems~\cite{tu2013speedy,dragojevic2015farm,kemper2011hyper}.

However, OCC is susceptible to aborts under contention. Most OCC mechanisms
perform read-set validation at the very end of a transaction. Any conflict
discovered there will cause the entire transaction to abort and restart, causing
wasted work. This
is particularly a problem for long-running read-heavy transactions~\cite{huang1991experimental}.

A number of systems have been proposed to address OCC's performance issues
under high contention. For example, \emph{contention managers} can prioritize
the execution of certain transactions over others to ensure better forward
progress~\cite{guerraoui2005cm,dragojevic2009swisstm}. \emph{Mixed} or \emph{adaptive concurrency control}
can dynamically decide to use pessimistic techniques, such as two-phase
locking, on frequently-written data~\cite{tang2017adaptivecc,narula14phase}. And OCC-based
algorithms have been proposed that can commit strictly more transactions than
conventional OCC. For example, TicToc can commit transactions that OCC would
normally abort, without violating serializability, by tracking two
timestamps per data item (a ``most recently read'' timestamp is added to conventional OCC's ``most recently written'' timestamp)~\cite{yu2016tictoc}.
The papers introducing these ideas have shown significantly reduced aborts and
better performance than OCC under high contention benchmarks.

We set out to replicate these results to validate the effectiveness of the
proposed optimizations. We implemented one variant for each of
the three classes of OCC-improving mechanisms mentioned above: SwissTM's contention manager~\cite{dragojevic2009swisstm}, a mixed concurrency control scheme based on our own adaptive reader-writer lock design, and TicToc.
We implement all these systems and conduct our experiments on top of
STO~\cite{herman2016sto}. We were expecting to see
these new mechanisms outperform the default OCC-based transaction engine in
STO for high-contention workloads.

But that is not what we saw. Careful scrutiny
of both the CC mechanisms in question and the workload itself
led to this observation: in one of the most widely-used
benchmarks, TPC-C~\cite{tpcc}, \emph{coarse-grained timestamps} are largely to blame
for OCC's poor performance at high contention. By using fine-grained
timestamps assigned to multi-column values, OCC outperforms all three
OCC-improving mechanisms mentioned above. By reporting our
findings, we show that the granularity of a CC mechanism is a crucial
performance parameter not to be overlooked in the context of modern in-memory
database systems.

%% file: related.tex
\section{Related Work}
\label{rw}

TDSL, the Transactional Data Structure Library~\cite{tdsl}, and
STO, Software Transactional Objects~\cite{herman2016sto}, are STM frameworks integrated with libraries of
transaction-aware data types. By testing conflicts at a higher level than
previous word-based STMs~\cite{dice2006tl2}, these systems can reduce the
overheads associated with large transaction tracking sets and the frequency of
false conflicts. When used as an in-memory database, STO can outperform
previous purpose-built database systems~\cite{tu2013speedy}.
TDSL evaluates STM benchmarks, but offers support for extremely efficient
single-operation transactions.
We implement our in-memory database using STO, and extend STO to support
multiple concurrency control mechanisms.

Contention managers~\cite{guerraoui2005cm}
are enhancements to OCC addressing the issue
of aborts under high contention. Instead of simply aborting
and re-executing upon observing a conflict, the transactional system
consults the contention manager to decide which transaction to abort.
The contention manager makes the decision by assigning different priorities to
different transactions, with the goal of ensuring the forward progress of the
overall system.
SwissTM~\cite{dragojevic2009swisstm} is an example of an STM system
that has a contention manager built in. It promises $\sim1.16\times$ speed-up
over TL2~\cite{dice2006tl2} and can achieve comparable performance to
more advanced type-based TM systems.
%
We implement our own version of the SwissTM contention manager on top of STO and examine how it performs under a database workload.

There has been recent interest in systems that use mixed modes of
concurrency control. Adaptive Concurrency Control
(ACC)~\cite{tang2017adaptivecc} is a system that dynamically
switches between pessimistic and optimistic CC based on a
pre-defined set of workload features like conflict rate, read/write ratio, etc.
The system works in a partitioned fashion, where each
CC mechanism operates exclusively in one partition, and the system
ships data between partitions to select the appropriate CC
mechanism for each record. The resulting mechanism is shown to
outperform OCC under contended workloads.
We implement our own adaptive CC mechanism based on a reader-writer lock. To
lower the overhead of the mechanism, we do not use a partitioned approach, but
create a unified commit protocol instead that handles different CC policies
executed in the same transaction. We use a state machine per record to
regulate the CC mechanism in use. Our approach is more lightweight compared to
many other adaptive mechanisms and therefore better suits single-node
in-memory database systems.

TicToc~\cite{yu2016tictoc} attempts to improve OCC from within. It uses
separate read and write timestamps for reach record, allowing for more
flexible transaction schedule reconciliation at commit time. TicToc is shown
to allow serializable transaction schedules not possible under OCC or 2-phase
locking. For example, consider the transaction
interleaving in~\autoref{fig:tictoc_interleaving}, where 2 transactions
access a table concurrently.
TicToc can commit both transactions in this scenario, but OCC or locking-based
mechanisms would have to abort at least one transaction. TicToc achieves this
by rescheduling Txn 1 to commit before Txn 2 in the serialization order, despite
Txn 1 finishing \emph{after} Txn 2 in real time.

\begin{figure}
  \small
  \centering
  \begin{tabular}{p{1in}l}
  \textbf{Txn 1}                & \textbf{Txn 2}                   \\
  \multirow{2}{*}{read row A}   &                                  \\
								& \multirow{2}{*}{update row A}    \\
  \multirow{2}{*}{update row B}                                    \\
								& \multirow{2}{*}{commit}          \\
  \multirow{2}{*}{commit}                                          \\
  \end{tabular}
  \caption{Example of a transaction interleaving where TicToc can commit both
  transactions but OCC can not.}
  \label{fig:tictoc_interleaving}
\end{figure}


A potential issue with TicToc
is read timestamp maintenance: atomic compare-and-swap
operations may be issued to shared
metadata of objects that are merely read during the
transaction. This appears to undermine a major performance argument of OCC. We
examine this further by implementing our own version of TicToc.

Early work on database management systems showed that
locking granularity impacts performance~\cite{ries1977effects}.
This suggested that finer granularity
doesn't always lead to better performance, but having {\em some} level of
fine-grained locks is still better than having just one coarse-grained global
lock. Despite its age, the study's conclusion appears to hold even
today. For example, SwissTM finds that
4 STM words per lock achieves better performance than 1 word per lock~\cite{dragojevic2009swisstm}, and coarse-grained locking can 
outperform fine-grained locking in some
Java-based benchmarks by as much as 3$\times$~\cite{govassilis2014techblog}.

Much work on synchronization granularity concerns pessimistic locking,
rather than the OCC-based synchronization widely used in modern in-memory
database systems. Our work shows that fine-grained timestamps in OCC can be a
clear win in real-world benchmarks, largely because of OCC's low overhead per
timestamp. We show that the gains achieved by
using fine-grained timestamps in OCC are greater than those achieved by
using the aforementioned alternative CC mechanisms.

%% file: impl.tex
\section{Implementation}
\label{impl}

This section is an overview
of the software platform we use to obtain our results.

\subsection{STO}

We build an in-memory database in STO by constructing a special
datatype for database indexes and tables. The datatype we implemented is based
on Masstree~\cite{mao2012masstree}, a cache-friendly concurrent B-tree data
structure. We extended Masstree under the STO framework to support transactional
insert, select, update, scan, and delete operations.

\subsection{Flexible Concurrency Control}

We extended STO to support other CC
schemes, specifically
Adaptive Read/Write Locking, 2-Phase Locking, SwissTM Concurrency Control, and
TicToc.

\paragraph{Adaptive Read-Write Locking} This mechanism uses an adaptive
reader-writer lock guarding access to each record. The reader-writer lock
automatically switches between optimistic mode (where reads don't acquire
locks but only observe versions, like in OCC) and pessimistic mode (where a
strict reader-writer lock is enforced) based on the level of contention
observed with the associated record.

\paragraph{2-Phase Locking} This is just the simple 2-Phase Locking mechanism
where a lock has to be acquired before accessing a record. We use reader-writer
locks.

\paragraph{SwissTM Concurrency Control} In this mechanism, we use SwissTM's
combination of an eager locking (for writes) OCC with a timestamp-based
contention manager.
The contention manager favors the longer-running
transaction based on the timestamps, and aborts the other transaction.

\paragraph{TicToc} We implemented TicToc as described in Yu et al.~\cite{yu2016tictoc},
modified according to their published implementation~\cite{github-dbx1000}.
Yu et al. describe several optimizations to the basic protocol. One
of these, non-waiting deadlock prevention (\S5.1 of~\cite{yu2016tictoc}), is enabled by default
in STO for all CC. Another, compressed 64-bit timestamps (\S3.6),
performed far worse than a simpler 128-bit implementation
because version number overflow caused TicToc to abort more than OCC.
We found that preemptive aborts (\S5.2) did not improve performance.

\paragraph{}
We designed and implemented a unified commit protocol providing simultaneous support
for these CC mechanisms within the same transaction.
This approach is correct as it only allows the intersection of the permitted
schedules of participating CC mechanisms.
%
The unified commit protocol implements OCC no slower than the
original OCC-only protocol.

\subsection{Benchmarks}

We evaluate these CC mechanisms using standard benchmarks, specifically
YCSB~\cite{cooper2010ycsb} and TPC-C~\cite{tpcc}.

Our YCSB-like workload uses the YCSB key-value schema, but groups several
operations into transactions. The YCSB table is prepopulated with 10 million
keys before the benchmark starts, each associated with a value that consists
of 10 columns (10 bytes per column). Each transaction comprises 16 operations,
$\sim50\%$ reads and $\sim50\%$ writes, which randomly selects a key and
randomly reads or writes one of its associated columns. To simulate a
high-contention workload, transactions access keys according to a Zipfian
distribution with $\theta=0.9$ over all keys in the database.

TPC-C is a widely-used industry standard evaluating the performance of
transaction processing systems. It models a moderately-complex transactional
inventory management application. We implement the New-order, Payment, and
Order-status transactions of TPC-C, which account for 92\% of all transactions
in the TPC-C default workload mix. The contention level of TPC-C can be
controlled by varying the number of ``warehouses''; fewer warehouses per
core means more contention.

As is typical, we associate each value (i.e., each 10-column YCSB value and
each TPC-C row) with a single timestamp. Any change to a value updates the
timestamp, invalidating any concurrently-reading transactions.

\subsection{Timestamp Granularity}

TicToc claims to outperform OCC on TPC-C and other benchmarks by reordering
transactions, as shown in \autoref{fig:tictoc_interleaving}. We analyzed the
TPC-C benchmark to understand where this reordering occurred. Due to space
constraints we describe only the instance involving District tables:

\begin{compactitem}
\item A New-order transaction in TPC-C reads a
row from a District table to access the district tax rate information. This is
a read-only operation.

\item Concurrent Payment transactions may update the district
year-to-date amount (YTD) field in the same District table row, overwriting
the row that's read by a New-order transaction.

\item New-order and Payment transactions subsequently write to completely
different tables.
\end{compactitem}

\noindent
In this case the New-order transaction acts like Txn~1 and the Payment
transaction acts like Txn~2 in \autoref{fig:tictoc_interleaving}.

%
%
%

Note, though, that New-order and Payment transactions operate on disjoint
fields of the same row. If the field updated by Payment had a different
timestamp than the rest of the row, OCC would see no conflict at all. The
conflict it does detect is actually false.

TicToc's multi-part timestamp manages to partially mitigate this false
conflict based on other properties of the transaction (namely the relative
orders of writes and reads within each transaction), but it is also possible
to address the false conflict directly. We implement \emph{fine-grained
timestamps} in TPC-C by associating each District and Customer row with two
timestamps, one guarding the rarely updated fields of the row (e.g tax rate
for District tables) and another guarding the rest. In YCSB, we implement one
timestamp for even-numbered columns and one for odd-numbered columns. Locking
mechanisms use fine-grained locks rather than timestamps.

Our main evaluation questions are whether fine-grained timestamps have
significant overhead, and if not, whether they allow conventional OCC to
perform on par with other CC schemes.

%% file: eval.tex
\begin{figure*}[t!]
  \centering
  \begin{subfigure}[t]{0.48\textwidth}
      \centering
      \includegraphics[width=\textwidth]{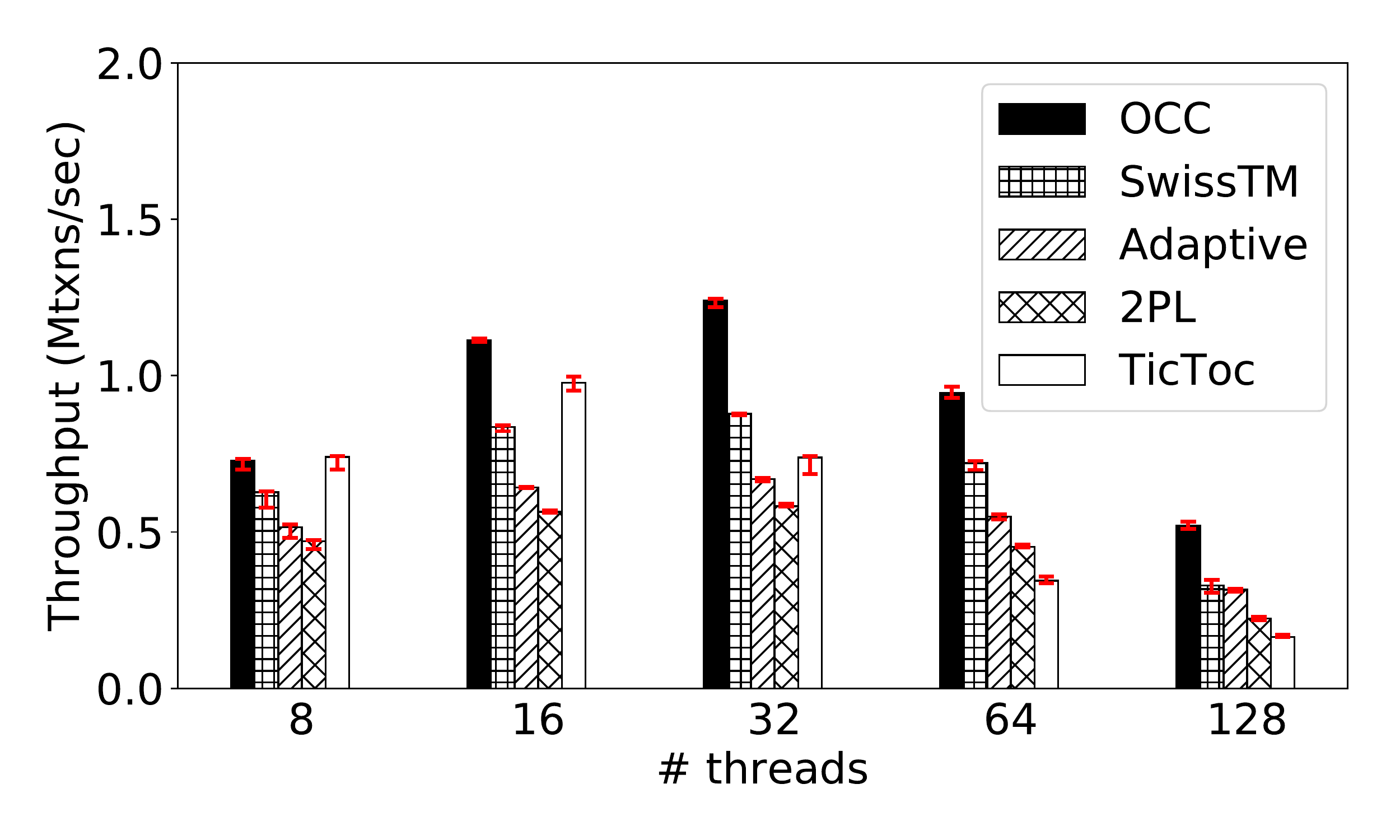}
      \caption{Coarse-grained timestamps (one timestamp for all columns).}
      \label{fig:ycsb_coarse_high}
  \end{subfigure}%
  \hspace{0.03\textwidth}
  \begin{subfigure}[t]{0.48\textwidth}
      \centering
      \includegraphics[width=\textwidth]{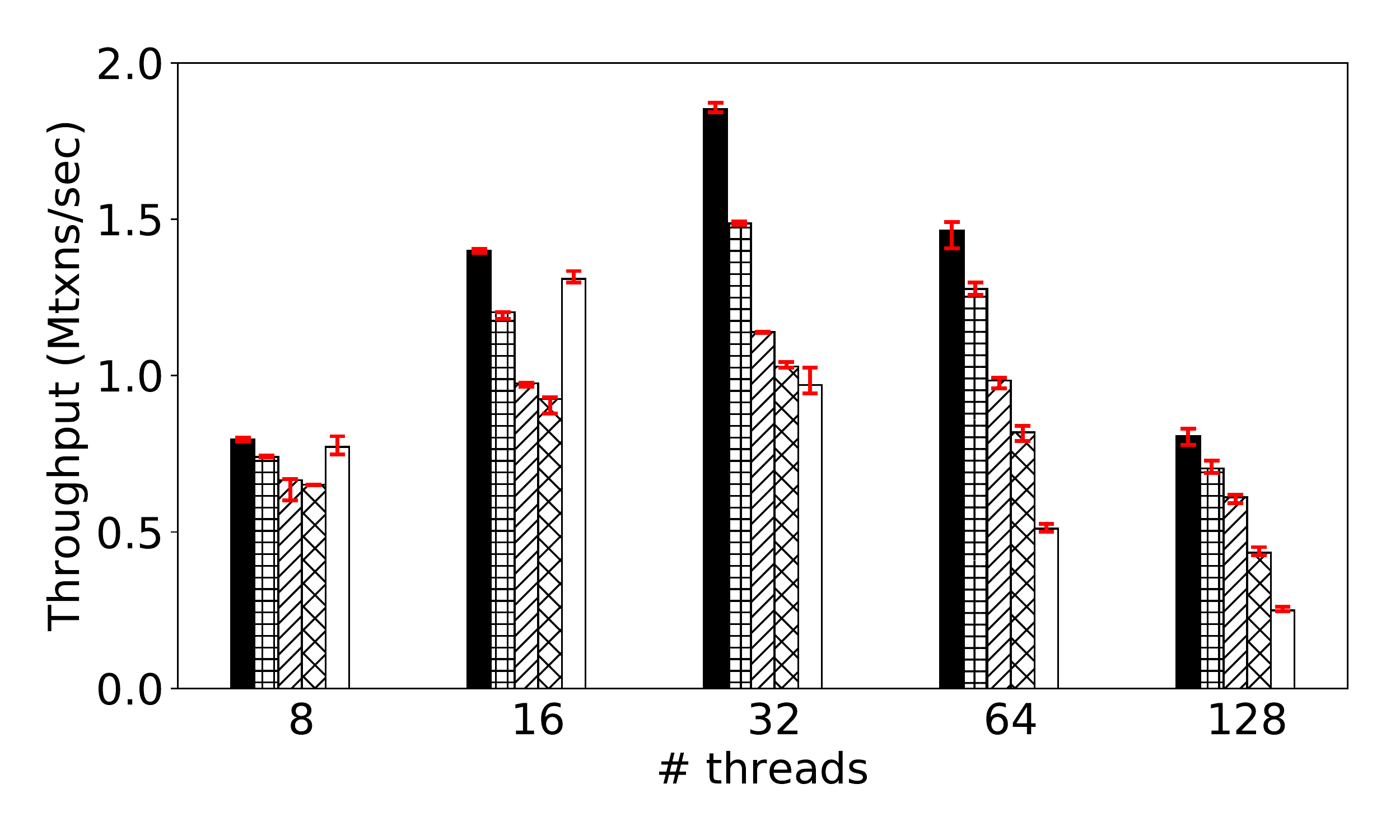}
      \caption{Fine-grained timestamps (one timestamp for odd-numbered columns,
        a second timestamp for even-numbered ones).}
      \label{fig:ycsb_fine_high}
  \end{subfigure}

  \caption{Throughput of YCSB-like workload with high contention (50\% writes, Zipfian $\theta=0.9$).}
  \label{fig:ycsb_high}

\end{figure*}

\begin{figure*}[t!]
  \centering
  \begin{subfigure}[t]{0.48\textwidth}
      \centering
      \includegraphics[width=\textwidth]{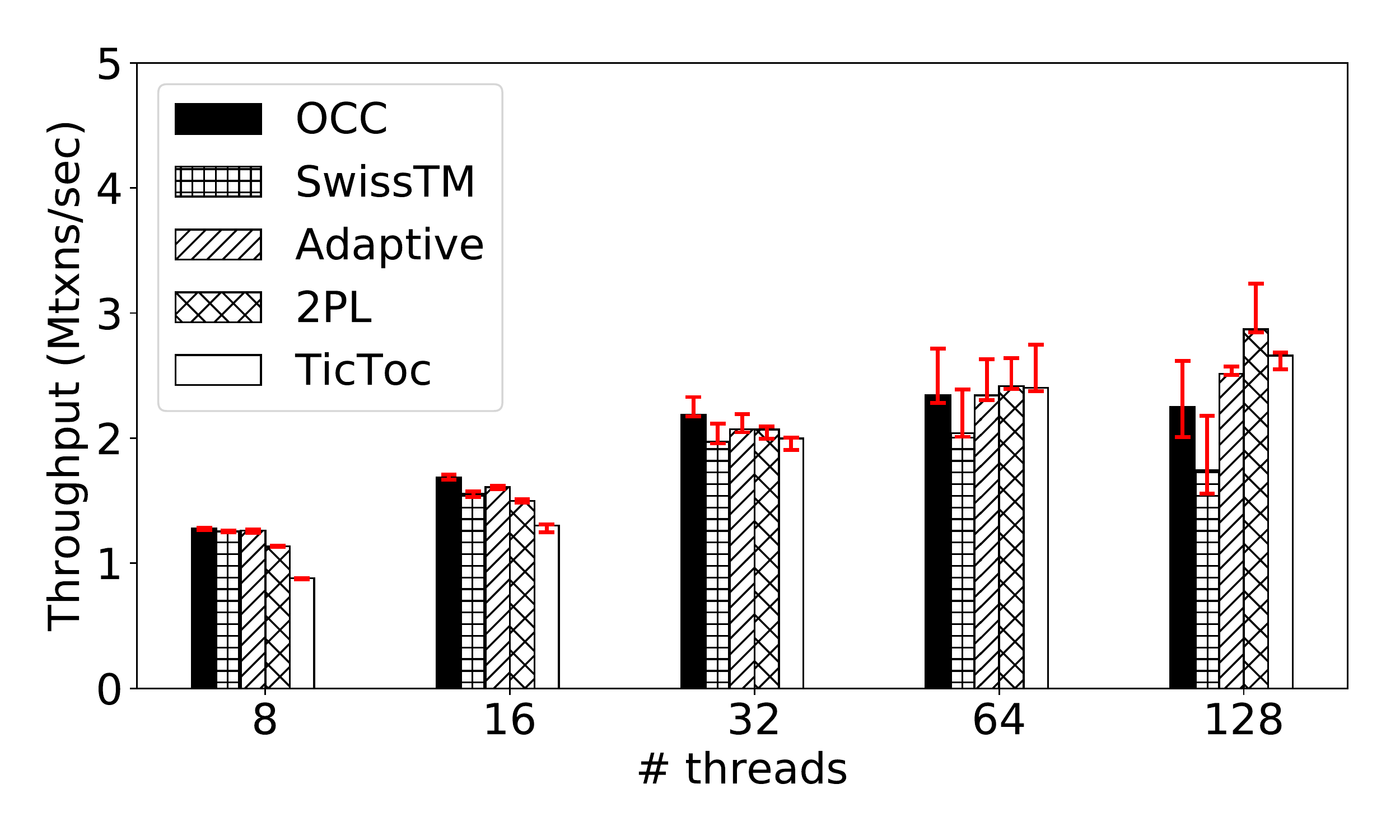}
      \caption{Coarse-grained timestamps (one timestamp per row).}
      \label{fig:tpcc_coarse_high}
  \end{subfigure}%
  \hspace{0.03\textwidth}
  \begin{subfigure}[t]{0.48\textwidth}
      \centering
      \includegraphics[width=\textwidth]{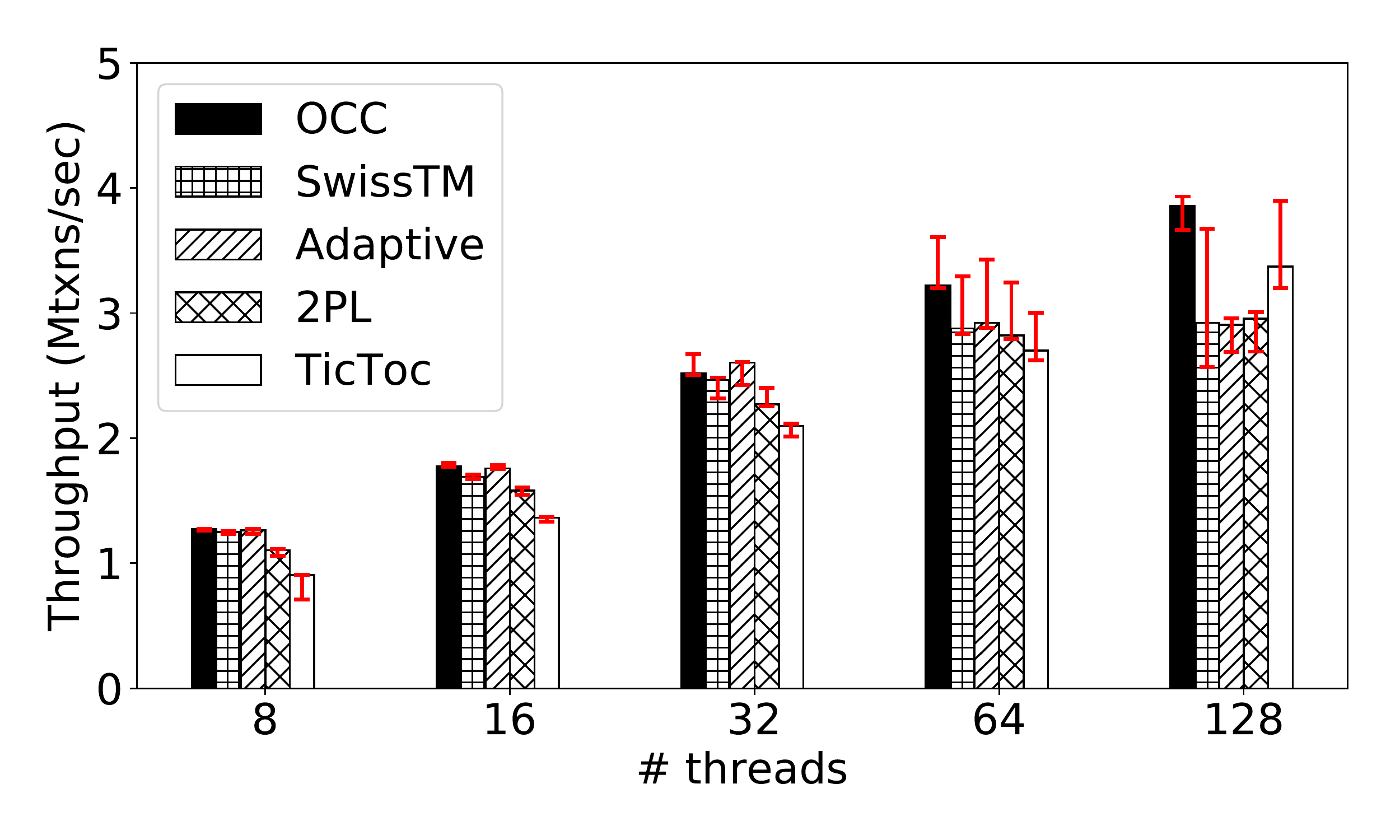}
      \caption{Fine-grained timestamps (two timestamps per row).}
      \label{fig:tpcc_fine_high}
  \end{subfigure}

  \caption{TPC-C throughput with increasing contention (8 warehouses fixed).}
  \label{fig:tpcc_high}
\end{figure*}

\section{Experiments}
\label{exp}

In this section, we describe our experimental setup and results of the TPC-C 
and YCSB benchmarks.

\subsection{Experimental Setup}

We run our experiments on a machine with 8 Intel Xeon E7-8990 v4 CPUs @ 2.20 GHz. The system has a total of 192
physical cores and 512 GB of RAM. Each experiment runs for a fixed
time duration of 15 seconds, and we report the median of 7 consecutive runs,
with maxes and mins shown as error bars.

\subsection{YCSB-like Workload}

The YCSB-inspired synthetic microbenchmark
evaluates these CC systems under high-contention stress.
(At lower contention levels, as expected, all mechanisms perform well.)
%
Results of the experiments are shown in \autoref{fig:ycsb_high}.
Because the performance trend closely tracks that
described in Yu et al.~\cite{yu2016tictoc}, and because our implementation of
TicToc outperforms in absolute terms the TicToc implementation reported
in its paper, we believe we have TicToc implemented fairly.\footnote{After
64 threads, the absolute performance of our implementation is slightly below
that in Yu et al. We believe this is because our machine has twice the
number of sockets and a more complex topology than theirs.}

\autoref{fig:ycsb_coarse_high} shows the results of the benchmark when we use
a coarse-grained timestamp for each value.
As the number of
threads increases, increasing conflict rates lead to performance
degradation for all CC mechanisms. TicToc starts off better than
OCC due to its ability to reschedule transactions and avoid aborts, but ends
up much worse than all other CC mechanisms as parallelism increases.
The overhead of extending TicToc read timestamps
via compare-and-swap, especially under a high contention workload,
is pronounced on this many-core machine, leading to more aborts.
The other mechanisms (SwissTM, Adaptive, and 2PL) perform uniformly worse than
OCC. These mechanisms were designed to provide benefits for specific
transaction mixes; for example, SwissTM ensures forward progress for
long-running transactions. But many mixes are not favorable, and unfavorable
mixes instead highlight the mechanisms' overheads.

\autoref{fig:ycsb_fine_high} uses finer-grained timestamps/locks, which
reduce, but do not eliminate, conflicts. All CC mechanisms benefit from
increased timestamp granularity in this benchmark, with OCC and SwissTM
observing the most gains compared to coarse-grained timestamps.

\subsection{TPC-C Workload}

The TPC-C workload is a more realistic transaction-processing workload than
YCSB. To model high contention, our TPC-C configuration fixes the number of
warehouses at 8,\footnote{8 warehouses matches the number of NUMA nodes in our
system.} so conflicts will occur due to concurrent accesses to the same
warehouse as the core count increases.

\autoref{fig:tpcc_coarse_high} shows TPC-C results with coarse-grained
timestamps. As in previous work, TicToc achieves gains over other CC
mechanisms as contention increases, up to 96 threads (data point not shown).
At 64 threads, TicToc achieves an abort rate of only 9.79\%,
significantly lower than that of the closest performing
system in terms of aborts, OCC, which reports an abort rate of 17.57\%.
TicToc's performance
begins to degrade at 128 threads, losing to 2 Phase Locking. This performance
drop at high core counts is consistent with the results shown in the TicToc
paper, and the trend is also observed, as expected, with all other optimistic
mechanisms (OCC and SwissTM). The absolute performance difference between
TicToc and OCC is less than the maximum $1.8\times$ previously reported,
however, due most likely to STO's baseline OCC implementation being more
efficient than that on TicToc's measurement platform, DBx1000~\cite{yu2014dbx1000,github-dbx1000}
(e.g., STO defaults to non-waiting deadlock
prevention).

\autoref{fig:tpcc_fine_high} introduces fine-grained timestamps/locks on
District/Customer tables, allowing all mechanisms to avoid false conflicts.
Unlike with coarse-grained timestamps, OCC is now the fastest mechanism
\emph{at almost all measured core counts} (Adaptive has a slight advantage at
32 cores). Switching to fine-grained timestamps reduces OCC's abort rate at
128 threads from a whopping 30.91\% to 1.75\%, the largest
drop observed in all systems.
It appears that, at least on TPC-C, TicToc's
benefits are due to false conflict avoidance, not true conflict rescheduling.
Similar to YCSB, all mechanisms increase in throughput.
TicToc never out-performs OCC but still manages to gain
ground over locking-based mechanisms, showing that fine-grained timestamps
improve the performance of optimistic mechanisms more than
pessimistic ones.

In these experiments we do not observe any slowdowns due to increased timestamp
granularity. The benefits of avoiding false conflicts in this benchmark greatly
outweigh the overhead incurred by maintaining more timestamps per record.
Furthermore, these results demonstrate that TicToc and other mechanisms'
performance benefits at high core counts in \autoref{fig:tpcc_coarse_high}
are also achievable by using OCC with fine-grained timestamps. In fact,
OCC with fine-grained timestamps outperforms TicToc with coarse-grained
timestamps by $1.37\times$ at 96 threads, where TicToc's performance peaked.
When we use fine-grained timestamps for all systems, OCC still outperforms
TicToc by as much as $1.14\times$ at 128 threads.

%% file: conclusion.tex
\section{Future Work}

Our results open up doors to further investigations of the impact of
timestamp granularity on CC mechanisms. We would like to investigate whether
this result applies to additional real-world
workloads.
TicToc can reschedule transactions with true conflicts; perhaps other
real-world workloads offer examples of important reschedulable true conflicts,
although we have found none yet.
We would
be interested in designing a CC scheme that can automatically detect false
conflicts due to coarse-grained timestamps and address them by dynamically increasing
timestamp granularity. We also plan to perform similar evaluations
on distributed CC mechanisms to test our findings in a distributed setting.

\section{Conclusion}

Aborts in OCC under high contention is a focus of research in in-memory
concurrency control, and various CC mechanisms are proposed to address this
issue. In this work, we set out to verify the claims made by various OCC improvements.
We implement and measure the performance of a variety of CC mechanisms using both
synthetic and real-world high contention workloads. We demonstrate that
timestamp granularity plays a significant role in the performance of all CC
mechanisms. In particular, the improvements to OCC by the alternative CC
mechanisms we implemented are also achievable by baseline OCC when fine-grained
timestamps are used, and OCC out-performs all other CC mechanisms when
fine-grained timestamps are in use at high core counts. Our findings demonstrate
that timestamp granularity has a greater impact on the performance of CC
mechanisms than previously thought, and it has been somewhat overlooked
during recent quest for faster or more complicated CC mechanisms.
We plan to extend our evaluation to more workloads in the future to
further solidify our claim that timestamp granularity should be treated
as a complementary avenue to addressing OCC's abort issue at high contention.
